\documentclass[aps,prl,reprint,superscriptaddress]{revtex4-2}

\usepackage[utf8]{inputenc}
\usepackage[T1]{fontenc}

\usepackage{newtxtext,newtxmath}

\usepackage[nolist]{acronym}

\usepackage{graphicx}
\usepackage{import}
\usepackage{transparent}
\usepackage{xspace}
\usepackage[table]{xcolor}
\usepackage{hyperref}
\hypersetup{
    colorlinks = true,
    linkcolor  = [rgb]{0.0,0.25,0.55},
    citecolor  = [rgb]{0.0,0.35,0.65},
    urlcolor   = [rgb]{0.0,0.3,0.6},
}
\usepackage{orcidlink}
\usepackage{cleveref}
\usepackage{array}
\usepackage{placeins}
\usepackage{amsmath}
\usepackage{amsfonts}
\usepackage{mathtools}
\usepackage{bm}
\usepackage[ruled,vlined]{algorithm2e}
\SetKwComment{Comment}{/* }{ */}
\usepackage[%
    print-unity-mantissa=false,
    range-phrase=--,
    range-units=single,
]{siunitx}

\graphicspath{{./}}

\newcommand*\diff{\mathop{}\!\mathrm{d}}
\newcommand{\thetaEOS}{\theta_{\rm{EOS}}}

\newcommand{\Msun}{{\rm{M}}_{\odot}}
\newcommand{\nbreak}{n_{\rm{break}}}
\newcommand{\nsat}{n_{\rm{sat}}}

\newcommand{\GRASP}{\affiliation{Institute for Gravitational and Subatomic Physics (GRASP), Utrecht University, Princetonplein 1, 3584 CC Utrecht, The Netherlands}}
\newcommand{\Nikhef}{\affiliation{Nikhef, Science Park 105, 1098 XG Amsterdam, The Netherlands}}
\newcommand{\Potsdam}{\affiliation{University of Potsdam, Institute of Physics and Astronomy, Karl-Liebknecht-Str. 24/25, 14476 Potsdam, Germany}}
\newcommand{\AEI}{\affiliation{Max Planck Institute for Gravitational Physics (Albert Einstein Institute), Am M\"{u}hlenberg 1, 14476 Potsdam, Germany}}

\begin{acronym}
\acro{ET}[ET]{Einstein Telescope}
\acro{MCMC}[MCMC]{Markov chain Monte Carlo}
\acro{GW}[GW]{gravitational wave}
\acro{SMC}[SMC]{sequential Monte Carlo}
\acro{NF}[NF]{normalizing flow}
\acro{ESS}[ESS]{effective sample size}
\acro{NS}[NS]{neutron star}
\acro{BNS}[BNS]{binary neutron star}
\acro{EOS}[EOS]{equation of state}
\acro{EFT}[EFT]{effective field theory}
\acro{MM}[MM]{metamodel}
\acro{TOV}[TOV]{Tolman-Oppenheimer-Volkoff}
\acro{SNR}[SNR]{signal-to-noise ratio}
\acro{NICER}[NICER]{Neutron star Interior Composition ExploreR}
\end{acronym}

\begin{document}

\title{Scalable and sequential inference of the neutron star equation of state with the Einstein Telescope}

\author{Thibeau~Wouters\,\orcidlink{0009-0006-2797-3808}}
\email{t.r.i.wouters@uu.nl}
\GRASP\Nikhef

\author{Thomas~C.~K.~Ng\,\orcidlink{0000-0002-9491-1598}}
\Nikhef\GRASP

\author{Hauke~Koehn\,\orcidlink{0009-0001-5350-7468}}
\Potsdam

\author{Fabian~Gittins\,\orcidlink{0000-0002-9439-7701}}
\GRASP\Nikhef

\author{Peter~T.~H.~Pang\,\orcidlink{0000-0001-7041-3239}}
\Nikhef\GRASP

\author{Tim~Dietrich\,\orcidlink{0000-0003-2374-307X}}
\Potsdam\AEI

\author{Chris~Van~Den~Broeck\,\orcidlink{0000-0001-6800-4006}}
\GRASP\Nikhef

\date{\today}

\begin{abstract}
Future gravitational-wave observatories such as the Einstein Telescope will detect tens of thousands of binary neutron star mergers per year, making gravitational waves a dominant probe of the neutron-star equation of state in the coming decades.
However, extracting this information requires hierarchical inference across a rapidly growing catalog of events, and existing methods must restart from scratch whenever a new event arrives, making them impractical at this scale.
In this Letter, we introduce a hybrid sequential Monte Carlo algorithm that combines data and likelihood tempering to adaptively update the equation-of-state posterior in batches, reusing previous inference results instead of restarting from a wide prior.
Accelerated by GPU hardware, our method infers the equation of state from a simulated month of around $1500$ binary neutron star mergers in a few hours on a single GPU, and we project that a full year of detections could be processed in a few days on current hardware.
This establishes sequential Monte Carlo as a practical route for real-time hierarchical inference with future gravitational-wave detectors.
\end{abstract}

\acresetall


\maketitle

\pdfbookmark[1]{Introduction}{sec:introduction}
\emph{Introduction} --- \Acp{NS} are unique laboratories to study cold, strongly interacting matter~\cite{Lattimer:2004pg,Ozel:2016oaf,Burgio:2021vgk,Chatziioannou:2024jsr}, and observing their properties can constrain the \ac{EOS} of matter at supranuclear densities~\cite{Koehn:2024set}.
Besides radio timing~\cite{Hobbs:2006cd,Demorest:2010bx} and X-ray pulse profile modeling~\cite{gendreau2016neutron,Bogdanov:2021yip,Watts:2019lbs}, \acp{GW} emitted by \ac{BNS} mergers reveal the properties of this matter through an imprint left by the tidal deformations of the two \acp{NS}~\cite{Hinderer:2009ca,Damour:2009vw,Damour:2012yf}, as demonstrated for the first time by GW170817~\cite{LIGOScientific:2017vwq,LIGOScientific:2017ync,LIGOScientific:2017zic,LIGOScientific:2018cki,LIGOScientific:2019eut}.
The next generation of ground-based \ac{GW} observatories, the \ac{ET}~\cite{Punturo:2010zz, Hild:2010id} and Cosmic Explorer~\cite{Reitze:2019iox, Evans:2021gyd, Evans:2023euw}, will provide a leap forward for this science case, detecting up to tens of thousands of \ac{BNS} mergers per year~\cite{ET:2019dnz, ET:2025xjr}, several orders of magnitude more than the currently operational Advanced LIGO~\cite{LIGOScientific:2014pky}, Advanced Virgo~\cite{VIRGO:2014yos} and KAGRA~\cite{KAGRA:2020tym}.

This unprecedented volume of detections strains the current data analysis methods in the \ac{GW} community in two ways.
First, analyzing all \ac{GW} signals from separate mergers incurs a significant computational cost~\cite{Couvares:2021ajn,Hu:2024mvn}.
Second (and the focus of this Letter) is the challenge presented by hierarchical inference~\cite{Thrane:2018qnx,Mandel:2018mve}, where the information from individual sources is combined to distill insights into properties of the population.
This is a crucial step to understand the mass distributions of compact objects~\cite{LIGOScientific:2026ctl}, study cosmology~\cite{LIGOScientific:2026ifv}, test general relativity~\cite{LIGOScientific:2026oim}, and, as we focus on in this work, determine the \ac{NS} \ac{EOS}~\cite{DelPozzo:2013ala,Agathos:2015uaa}.
It is as yet unclear how hierarchical inference will be performed with next-generation \ac{GW} data, and developing these methods early is essential to enable extensive projection studies~\cite{Regimbau:2012ir,Branchesi:2023mws,Tania:2025bsa}.

To solve this data analysis problem, the community requires hierarchical inference methods that are both scalable and sequential. 
Scalability addresses the cost of hierarchical inference, which grows as more events are added to the catalog.
Recent works have shown the potential of GPU accelerators to achieve scalable hierarchical inference for populations~\cite{Talbot:2019okv,Edelman:2022ydv,Talbot:2024yqw,Qazalbash:2025gny,Mancarella:2025uat,Mould:2025dts,Leyde:2026hvm}, the \ac{NS} \ac{EOS}~\cite{Wouters:2025zju}, cosmology~\cite{Papadopoulos:2026puy,Leyde:2026bat}, or a combination of those~\cite{Leyde:2023iof,Mastrogiovanni:2023emh,Mastrogiovanni:2023zbw,Tagliazucchi:2025ofb,Koehn:2026nin}. 
Besides scalability, it is also crucial to use a sequential approach that reuses previous inference results and updates posteriors as soon as we acquire new data.
However, so far, only a few works have formulated full Bayesian hierarchical inference algorithms that can sequentially process new data in `real time', without having to restart from a wide prior distribution~\cite{Leyde:2023iof,Wolfe:2026dcq,Demasi:2026axj} (although recent developments successfully achieved it for single-event GW analyses~\cite{Prathaban:2024rmu,Williams:2026jqv,Prathaban:2026kft}).
Moreover, none have demonstrated the applicability of such an approach to the inference of the \ac{EOS} from \ac{NS} mergers in the context of future detectors like \ac{ET}.

In this work, we provide a recipe for scalable, sequential hierarchical inference, and use it to constrain the \ac{EOS} with \ac{BNS} mergers observed by \ac{ET}.
Our approach combines a sequential Monte Carlo algorithm that adaptively processes the observations in batches, with fast inference delivered by GPU accelerators.
We demonstrate that, with current hardware, we can run the hierarchical inference process on one month of \ac{ET} observations of \ac{BNS} mergers in a matter of hours, and we project that one year of observations would take a few days.

\pdfbookmark[1]{Scalable, sequential hierarchical inference}{sec:sequential-inference-1}
\emph{Scalable, sequential hierarchical inference} --- After the individual \ac{BNS} mergers have been detected and analyzed, our hierarchical inference constrains the \ac{EOS} from the posterior distributions of the component masses and tidal deformabilities.
In this work, we do this by directly sampling the \ac{EOS} parametrization from Ref.~\cite{Koehn:2024set}, combining a nucleonic metamodel and an agnostic speed-of-sound extension scheme.
Further details are given in the Supplemental Material.

This sampling problem is notoriously computationally expensive, largely because it requires repeatedly solving the \ac{TOV} equations~\cite{Tolman:1939jz,Oppenheimer:1939ne} to predict the \ac{NS} macroscopic properties for each \ac{EOS} draw.
Therefore, many previous works have resorted to machine-learning surrogates to reduce inference cost (e.g.,~\cite{Ferreira:2019bny,Thete:2022drz,Reed:2024urq,Somasundaram:2024ykk,DiClemente:2025pbl,Rose:2026cnm}), although this adds training costs and reduces flexibility in the choice of prior and \ac{EOS} parametrization.

Rather than approximating the solutions for the \ac{TOV} equations, our method uses GPU hardware acceleration~\cite{jax2018github,frostig2018compiling}.
Beyond a GPU-native \ac{TOV} solver~\cite{Wouters:2025zju,kidger2021on}, two further aspects of our software exploit GPU parallelism.
First, we use normalizing flows~\cite{Kobyzev:2019ydm,Papamakarios:2019fms}, a type of neural density estimator, to turn single-event posterior samples into tractable hyperlikelihoods that can be evaluated at arbitrary points in their parameter space~\cite{Ruhe:2022ddi,Adamcewicz:2026pai}.
These flows enable high throughput on GPUs, especially when evaluated at single precision, even for over a thousand events.
Second, we sample the posterior using \ac{SMC}~\cite{del2006sequential,chopin2020introduction,dai2022invitation,cabezas2024blackjax}, which bridges from the prior to the posterior through a sequence of intermediate densities. 
At each iteration $t$, a set of particles approximating the current density is propagated to the next one by reweighting, resampling, and moving the particles with a \ac{MCMC} kernel.
Each of these operations is highly parallel and therefore benefits from GPU acceleration.
One version of \ac{SMC} uses \textit{likelihood tempering} to construct the intermediate densities, by raising the likelihood to an inverse temperature $\beta_t$. 
This artificial temperature can be set adaptively by requiring that each iteration preserves a certain \ac{ESS}~\cite{kish_ess}, which measures the diversity in a set of weighted samples.
This choice has recently gained popularity in the \ac{GW} community for analyzing individual \ac{GW} events~\cite{Karamanis:2022alw,Karamanis:2022ksp,karamanis2025persistent,Williams:2025szm,Williams:2025aar,Demasi:2026ltw}, and we recently used it for hierarchical inference on multi-messenger \ac{BNS} signals observed by \ac{ET}~\cite{Koehn:2026nin}.

More importantly, \ac{SMC} is naturally suited for sequential hierarchical inference, where data points are added `in real time' and previous posteriors are reused, rather than rerunning the inference from scratch.
For this, we take the intermediate densities in \ac{SMC} to be the partial posteriors, i.e., the intermediate density at iteration $t$ is the posterior conditioned on the first $t$ data points~\cite{dai2022invitation}.
A transition between densities is therefore realized by adding an extra data point to the likelihood, and this version of \ac{SMC} is known as \textit{data tempering}~\cite{chopin2002sequential}.
Whereas likelihood tempering gradually introduces the full likelihood through an artificial temperature parameter, data tempering does so by gradually introducing individual data points.

\begin{figure}[tpb]
    \includegraphics[width=\columnwidth]{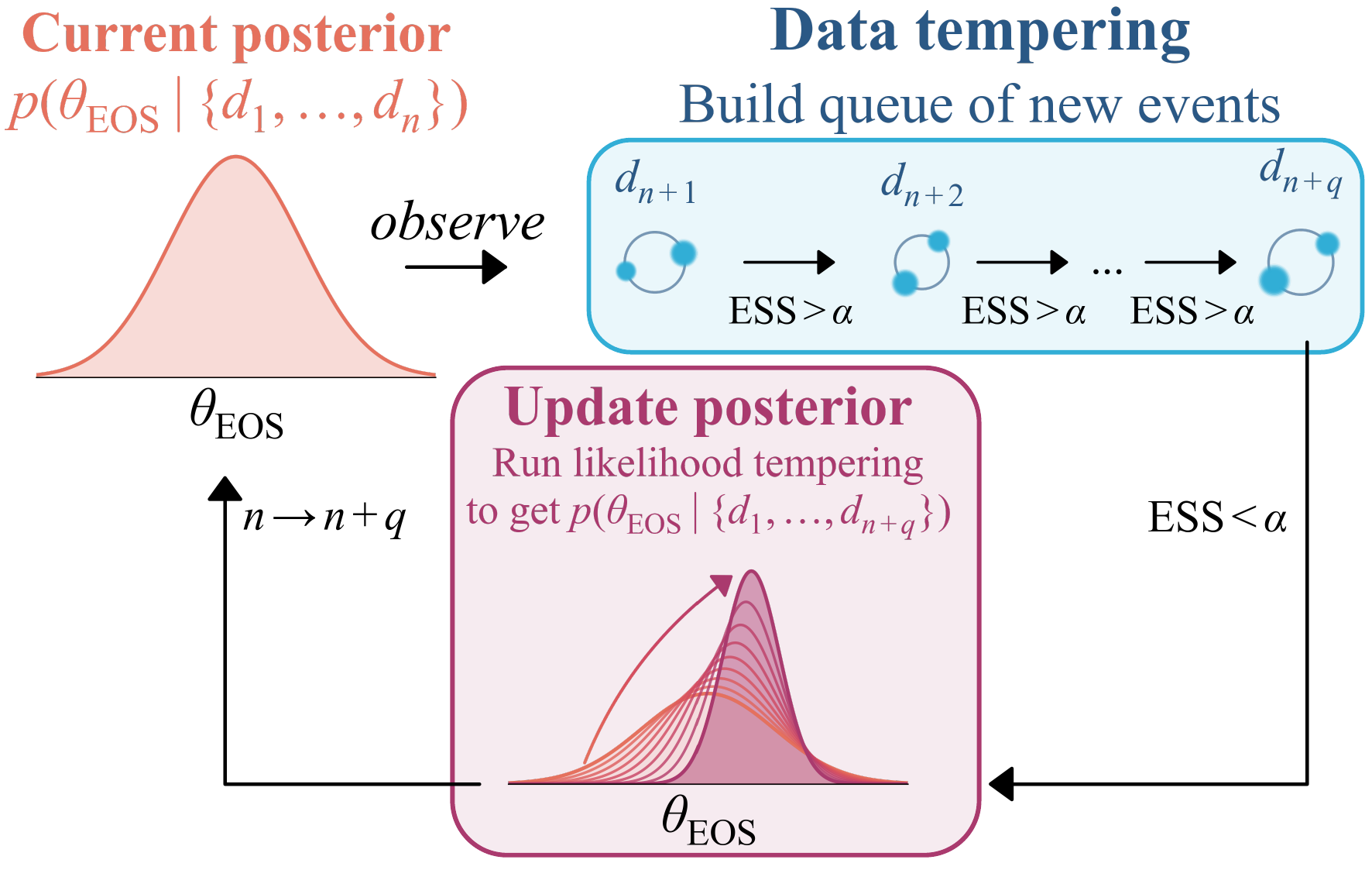}
    \caption{
        Illustration of our algorithm processing a growing catalog.
        Each iteration maintains a set of particles distributed according to the partial posterior conditioned on the $n$ events observed so far (\textit{upper left}).
        Observations that are not informative enough to significantly affect the posterior are collected in a queue (data tempering, \textit{upper right}).
        If, after accumulating $q$ events, the batch of new events is sufficiently informative, we update the partial posterior using likelihood tempering (\textit{lower}).
        Afterwards, we update our current posterior to be informed by $n+q$ events, and continue the loop.
        }
    \label{fig:diagram}
\end{figure}

However, a naive implementation of data tempering, where a transition adds one data point at a time, has a significant drawback. 
At first, reweighting particles with a highly informative data point can significantly reduce the \ac{ESS}, leading to low diversity, poor exploration of the parameter space, and ultimately biased posteriors. 
In this case, it would be better to perform additional substeps within a single iteration of data tempering.
However, at later stages, new observations might have little impact on the posterior when many events have already been incorporated, and such an update step may waste computational resources.
Taken together, we require a method to adaptively update posteriors with batches of new data points, where the batch size is determined based on the history of past events and the properties of new sources. 

Therefore, we implement a hybrid algorithm that combines data and likelihood tempering, illustrated in Fig.~\ref{fig:diagram}.
We provide mathematical details and pseudocode in the Supplemental Material.
At each iteration, we maintain a particle set distributed according to the partial posterior obtained from the first $n$ events.
When a new data point is considered, we reweight the particles with the corresponding single-event likelihood, and compute the \ac{ESS} of the weighted particle set. 
If the \ac{ESS} remains above a user-defined threshold $\alpha$, the event does not significantly impact the posterior and an update step is not warranted. 
In this case, we simply add the event to a queue of events. 
However, if the batch gets sufficiently informative after $q$ events (i.e., the \ac{ESS} drops below the threshold $\alpha$), we update the posterior: we run \ac{SMC} with likelihood tempering to bridge from the partial posterior conditioned on $n$ events to the partial posterior conditioned on all $n + q$ events observed so far. 
After sampling terminates, we update our set of particles, clear the queue, and are prepared to add new events.

\pdfbookmark[1]{Inferring one month of data}{sec:one-month-1}
\emph{Inferring one month of data}---To demonstrate our algorithm, we simulate one month of \ac{BNS} mergers observed by \ac{ET} in a $2$L-misaligned configuration~\cite{Branchesi:2023mws}, with one detector in the Meuse-Rhine euregion and the other in Sardinia, assuming a local merger rate of ${106.6~\mathrm{Gpc}^{-3}~\mathrm{yr}^{-1}}$ (falling within current constraints~\cite{LIGOScientific:2026ctl}, and evolved over redshift~\cite{Madau:2014bja}) and Gaussian detector noise. 
We draw \ac{NS} masses uniformly between $[1, 2]~\Msun$ and assign tidal deformabilities based on a fiducial \ac{EOS}, which we determine using existing constraints from the mass measurement of PSR~J0740+6620~\cite{Fonseca:2021wxt}, \ac{NS} mass-radius measurements of pulsars observed by NICER~\cite{gendreau2016neutron,Bogdanov:2021yip,Watts:2019lbs} (in particular, taking posterior samples from Refs.~\cite{Riley:2019yda,Miller:2019cac,Miller:2021qha,Salmi:2024aum,Choudhury:2024xbk,Mauviard:2025dmd}), and GW170817~\cite{LIGOScientific:2017vwq,LIGOScientific:2017ync,LIGOScientific:2017zic,LIGOScientific:2018cki,LIGOScientific:2019eut}: the Supplemental Material gives more details.

From this set, we consider a source confidently detected if its network \ac{SNR} is above $12$~\cite{KAGRA:2013rdx}, yielding $1~579$ sources, of which $1~555$ are retained after removing a few events whose parameter estimation did not converge. 
Their source parameters are determined with nested sampling~\cite{Skilling:2004pqw,Skilling:2006gxv,Speagle:2019ivv,sergey_koposov_2026_21417233,Ashton:2018jfp, Romero-Shaw:2020owr}. 
However, since realistic settings for single-event analysis would make this step computationally infeasible, we adopt simplified settings.
In particular, we use a lower frequency cutoff of $20$ Hz, analyze signals with the faster but less realistic \texttt{TaylorF2} waveform model~\cite{Buonanno:2009zt,Blanchet:2013haa}, and only sample masses and tidal deformabilities. 
Further details are given in the Supplemental Material.
Nevertheless, the hierarchical inference step makes no simplifications in how the single-event posteriors are incorporated.

We then use this catalog to infer the \ac{EOS} by running our proposed \ac{SMC} algorithm. 
We start from an initial \ac{EOS} posterior that is informed solely by the mass measurement of PSR~J0740+6620~\cite{Fonseca:2021wxt}.
The \ac{ESS} threshold to trigger an update step is set to $50\%$ of the total particle number, which is fixed at $2000$.

\begin{figure}[t]
    \includegraphics[width=\columnwidth]{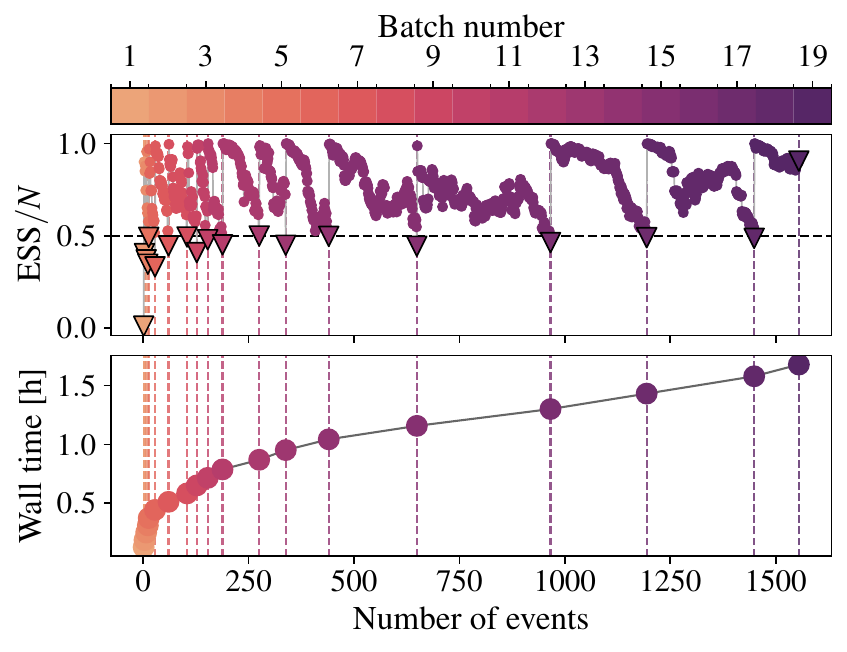}
    \caption{
        Diagnostic plots of our \ac{SMC} sampler.
        \textit{Top panel}: Sampling efficiency when reweighting the current set of posterior samples with respect to the new events. 
        The inverted triangles show iterations where the ESS falls below the $50\%$ threshold (dashed line), and we update the partial posterior with likelihood tempering.
        \textit{Bottom panel}: Cumulative wall time as a function of the number of events. 
        }
    \label{fig:smc-updating-diagnostics}
\end{figure}

Figure~\ref{fig:smc-updating-diagnostics} shows how the sampler processes events in batches, indicated by the color gradient.
The top panel shows the sampling efficiency (\ac{ESS} divided by total number of samples) after reweighting with each new event, while the bottom panel shows the cumulative wall time recorded after each update step (all analyses in this paper run on a single NVIDIA H100 GPU). 
Initially, we have little information on the \ac{EOS} and the first few events cause the ESS to drop quickly below the threshold value, despite most of these events having a moderate \ac{SNR} (our events are not ordered by \ac{SNR}, but instead based on their arrival times on Earth).
Moreover, different signals have sources at different masses, so that one detection in a previously unseen mass regime will be highly informative.
After about $100$ events, the posterior starts to stabilize, and updates occur less often.
At later stages, several hundred events can be added to the queue before an update step is triggered. 
In total, we require less than $2$ hours to process all events sequentially. 
In the Supplemental Material, we show that the same set of sources without noise in the detector leads to fewer update steps, showing that data quality can impact the runtime of our algorithm.

\begin{figure}[t]
    \includegraphics[width=\columnwidth]{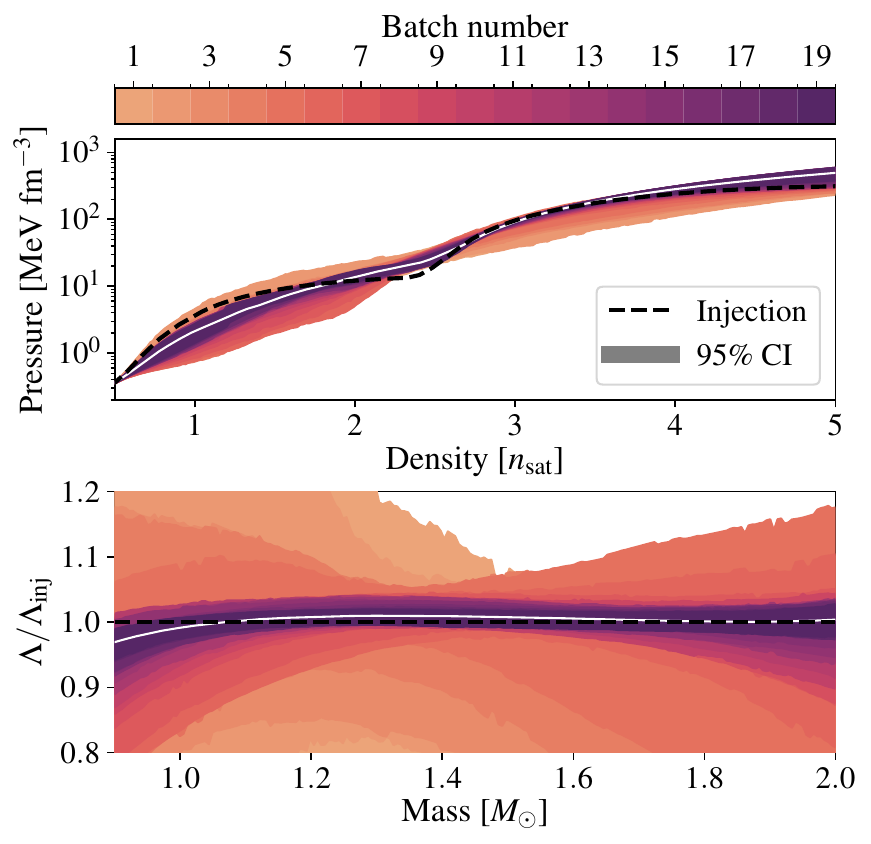}
    \caption{
        Constraints on the \ac{EOS} (top panel) and tidal deformability (bottom panel) after processing each batch, shown as $95\%$ credible bands.
        The injected \ac{EOS} is shown with a black dashed line, while the median of the posterior in the final batch (i.e., with all events processed) is shown in white.
        The median is taken at each value for the density or \ac{NS} mass.
    }
    \label{fig:eos-constraints}
\end{figure}

Figure~\ref{fig:eos-constraints} shows the constraints on the \ac{EOS} (top panel, with density in units of nuclear saturation density $\nsat\equiv 0.16~\mathrm{fm}^{-3}$) and the mass--tidal-deformability relationship (bottom panel) after each \ac{SMC} step.
The bulk of the posterior underestimates the pressure at lower densities, and overestimates it at intermediate densities. 
While the injected \ac{EOS} lies outside the $95\%$ credible band, this discrepancy reduces for a catalog without detector noise, as we show in the Supplemental Material. 
The recovery at the highest densities is less accurate, which is expected since our mock catalog is limited to \acp{NS} lighter than $2~\Msun$.
Despite these features, the fiducial tidal deformability curve is within the $95\%$ credible band for all batches, since the tidal deformability is an integral quantity over the \ac{EOS} and the observable that actually enters the likelihood.
In the final batch, we obtain constraints on the order of $1\%$ on the tidal deformability across the masses probed by our catalog. 
While the median slightly overestimates the tidal deformability at intermediate masses, this is again largely resolved when considering injections without detector noise.

These observations reinforce that the \ac{EOS} has degeneracies which are hard to resolve from mass--tidal-deformability measurements alone~\cite{Krastev:2018nwr,Mondal:2021vzt,Iacovelli:2023nbv,Wouters:2025zju,Zhu:2025dea}, and independent constraints are needed to break them.
This problem is worsened because our injection \ac{EOS} has a feature resembling a phase transition around $2.5 \, \nsat$~\cite{Raithel:2022efm,Raithel:2022aee}. 

\pdfbookmark[1]{Extrapolating to one year}{sec:one-year-1}
\emph{Extrapolating to one year}---To assess how well our algorithm scales with the number of events, we fit the runtime, recorded at each update step, as a function of the number of events, and extrapolate the fit. 
We find that the cost for each individual step is essentially flat up to $n \sim 400$ events, after which our calculations become memory-bound and the wall time increases roughly linearly with the number of events.
Therefore, we derive a power-law fit by only using data points with $n > 400$. 

We use this fit to predict the wall time required to analyze one month of more realistic \ac{BNS} detections made with \ac{ET}, in particular, starting the analysis from $5$ Hz and using the \texttt{IMRPhenomXAS\_NRTidalv3} waveform~\cite{Pratten:2020fqn,Abac:2023ujg}.
With these settings, we find that one month yields ${2~559}$ sources, and hence expect around ${30~000}$ detections after one year of observation. 
These numbers should be taken as rough estimates, as they depend on, e.g., detector performance~\cite{Negr:2026gwp}, the actual mass distributions of \acp{NS}~\cite{Landry:2021hvl}, and uncertainties in the merger rate~\cite{LIGOScientific:2026ctl}. 
The fit predicts that our hierarchical inference algorithm will take $2.4$ hours ($2.2$ days) to process one month (one year) of such more realistic observations on a single NVIDIA H100 GPU.

\begin{figure}[t]
    \includegraphics[width=\columnwidth]{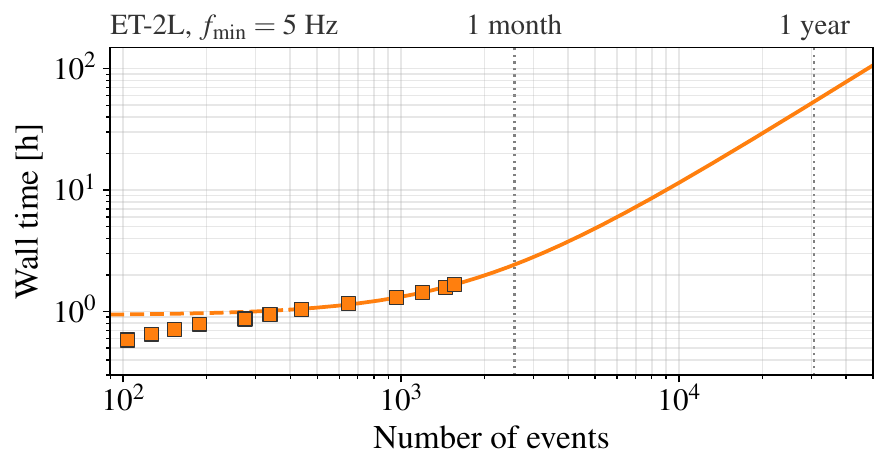}
    \caption{
        Extrapolation of the cumulative wall time as a function of the number of events.
        The markers show the runtimes recorded at the batches shown in Fig.~\ref{fig:smc-updating-diagnostics}, while the curve shows a fit based on the data points from $n>400$ onwards (solid line).
        The vertical lines show the expected number of detections for a catalog starting from a lower frequency of $5$ Hz, and analyzed with the \texttt{IMRPhenomXAS\_NRTidalv3} waveform.
    }
    \label{fig:extrapolations}
\end{figure}

\pdfbookmark[1]{Discussion and conclusion}{sec:conclusion}
\emph{Discussion and conclusion}---In this work, we present a scalable Bayesian hierarchical inference algorithm that can handle the large volume of detections enabled by future \ac{GW} instruments, such as \ac{ET}. 
Specifically, we use normalizing flows to stack single-event posteriors, accelerate likelihoods on GPUs, and parallelize sampling with \ac{SMC} samplers.
We also showed that \ac{SMC} is a natural framework to sequentially posteriors `in real time', by reusing past inference results and avoiding restarting analyses from scratch.
Our method determines adaptively when a batch of new events is sufficiently distinct from the current posterior and an update is warranted.
This allows us to rapidly analyze a growing catalog of events during observing runs as expected with \ac{ET}, where tens to hundreds of detections will be made every single day.

We demonstrated the performance of this algorithm by inferring the \ac{EOS} from one month of mock \ac{BNS} events within a few hours on a single GPU. 
Furthermore, we project that our \ac{SMC} sampler can infer the \ac{EOS} from a realistic full year of \ac{BNS} mergers within a few days.
The method presented here can be directly extended to perform joint inference on the \ac{EOS}, the \ac{NS} mass distribution, and cosmology, following our recent work~\cite{Koehn:2026nin},  
although this would require careful application of selection effects~\cite{Mandel:2018mve,Golomb:2021tll,Golomb:2024lds}.
Moreover, we believe our workflow to apply to other hierarchical inference problems as well, such as population studies of black holes, where sequential inference is already becoming crucial in current observing runs~\cite{Wolfe:2026dcq}. 

In the future, we will apply our algorithm to more realistic mock datasets, considering the full duration of the signal and using more recent waveform models.
Moreover, various effects can impact the \ac{EOS} inference, such as waveform systematics~\cite{Gamba:2020wgg,Kunert:2021hgm,Chatziioannou:2021tdi,Yelikar:2024wzm,Yelikar:2024rmh,Williams:2026jqv}, unmodeled physics in the waveform~\cite{Pratten:2021pro,Ghosh:2025glz,Gittins:2026ntx}, overlapping signals~\cite{Samajdar:2021egv,Antonelli:2021vwg,Janquart:2022fzz}, or potential noise artifacts~\cite{Pankow:2018qpo,Narikawa:2018yzt}.
With our scalable framework, we can investigate the robustness of \ac{EOS} inference under such influences using large catalogs.
Future work could also constrain the \ac{EOS} from neutron star-black hole mergers detected by future \ac{GW} observatories~\cite{Sarin:2023tgv,Dubey:2026pew}. 

Our algorithm could be optimized further in the following way. 
We can reduce the cost of the likelihood evaluation by approximating intermediate posteriors from already assimilated events with a normalizing flow, following Refs.~\cite{Wolfe:2026dcq,Williams:2025aar}. 
This flow would act as a surrogate density for previous partial posteriors, replacing the hyperlikelihood evaluation of already assimilated events, which otherwise dominates its cost.
Moreover, this would naturally allow us to perform \textit{model tempering}, where these posterior distributions can be reused in the analysis of an extended model, for instance, when adding parameters to account for \ac{NS} anisotropy~\cite{Pang:2025fes} or beyond-GR theories of gravity~\cite{Brown:2024joy}.
This idea has already proven to be successful in the \ac{GW} community for single-event analysis~\cite{Williams:2025aar}.

\emph{Data availability}---Our inference code, \textsc{jester}, is publicly available at~\cite{jester_github}. 
Our data products and scripts to run the \textsc{jester} inference are available on Zenodo~\cite{wouters_2026_22300363}.

\begin{acknowledgments}
\emph{Software used}---This work made use of the open-source software packages \textsc{dynesty}~\cite{Speagle:2019ivv,sergey_koposov_2026_21417233}, \textsc{bilby}~\cite{Ashton:2018jfp,Romero-Shaw:2020owr}, \textsc{gwmock}~\cite{gwmock_v0.13.2}, \textsc{jax}~\cite{jax2018github,frostig2018compiling}, \textsc{diffrax}~\cite{kidger2021on}, \textsc{flowjax}~\cite{ward2023flowjax}, \textsc{blackjax}~\cite{cabezas2024blackjax}, and \textsc{jester}~\cite{jester_github}.

\emph{Acknowledgments}---T.W. thanks Michael Williams for introducing him to \ac{SMC} samplers, and Vasudev Dubey, Sanika Khadkikar, and Praveer Tiwari for internal review of \textsc{jester} for the LIGO-Virgo-KAGRA collaboration. 
We thank Matthew Mould for valuable feedback during the internal LVK review of the paper.
T.W. and T.C.K.N. thank James Alvey, Metha Prathaban, and David Yallup for discussions regarding \textsc{blackjax} samplers.
T.W. and T.D. thank János Takátsy for discussions regarding figures.
T.W. and C.V.D.B. are supported by the research program of the Netherlands Organization for Scientific Research (NWO) under grant number OCENW.XL21.XL21.038.
P.T.H.P. is supported by the research program of the Netherlands Organization for Scientific Research (NWO) under grant number VI.Veni.232.021.
We thank SURF (www.surf.nl) for support in using the National Supercomputer Snellius under project numbers EINF-14622 and EINF-19565.
This material is based upon work supported by NSF's LIGO Laboratory which is a major facility fully funded by the National Science Foundation.
\end{acknowledgments}

\bibliography{references}

\FloatBarrier
\newpage
\onecolumngrid
\section{Supplemental Material}

\begin{acronym}
\acro{ET}[ET]{Einstein Telescope}
\acro{MCMC}[MCMC]{Markov chain Monte Carlo}
\acro{SMC}[SMC]{sequential Monte Carlo}
\acro{NF}[NF]{normalizing flow}
\acro{IBIS}[IBIS]{iterated batch importance sampling}
\acro{ESS}[ESS]{effective sample size}
\acro{GW}[GW]{gravitational wave}
\acro{NS}[NS]{neutron star}
\acro{BNS}[BNS]{binary neutron star}
\acro{EOS}[EOS]{equation of state}
\acro{MM}[MM]{metamodel}
\acro{TOV}[TOV]{Tolman-Oppenheimer-Volkoff}
\acro{SNR}[SNR]{signal-to-noise ratio}
\acro{NICER}[NICER]{Neutron star Interior Composition ExploreR}
\end{acronym}

\section{Hierarchical Bayesian inference of the equation of state}

The matter inside all neutron stars is assumed to follow a single, universal \ac{EOS}. 
One option to model it is through non-parametric methods, such as Gaussian processes~\cite{Landry:2018prl,Essick:2019ldf,Landry:2020vaw,Legred:2022pyp,Essick:2023fso,Finch:2025bao,Boudon:2025ztu,Ng:2025wdj,Suleiman:2025dkn}.
Here, we focus instead on sampling a flexible \ac{EOS} parametrization $\thetaEOS$ (described in more detail below). 
The hyperparameters $\thetaEOS$ are constrained by a collection of $N$ observations of \ac{BNS} mergers $\mathcal{D} = \{d_1, \dots, d_N\}$ through Bayes' theorem:
\begin{equation}
    \mathcal{P}(\thetaEOS|\mathcal{D}) = \frac{\mathcal{L}(\mathcal{D}|\thetaEOS) \pi(\thetaEOS)}{\mathcal{Z}} \, ,
\end{equation}
where $\mathcal{P}(\thetaEOS|\mathcal{D})$ is the hyperposterior, $\mathcal{L}(\mathcal{D}|\thetaEOS)$ the hyperlikelihood, $\pi(\thetaEOS)$ the prior and $\mathcal{Z}$ the Bayesian evidence.
When the observations are independent, the hyperlikelihood is given by~\cite{Thrane:2018qnx,Mandel:2018mve}
\begin{equation}
    \mathcal{L}(\mathcal{D}|\thetaEOS) = p_{\rm{det}}(\thetaEOS)^{-N} \prod_{i=1}^N \int \mathcal{L}\left( d_i| \theta_i\right) p\left( \theta_i|\thetaEOS \right) \diff \theta_i \, ,
\end{equation}
where $p_{\rm{det}}(\thetaEOS)$ encodes selection effects, and $\mathcal{L}\left( d_i|\theta_i \right)$ is the single-event likelihood.
In this study, our observed catalog is based on an \ac{SNR} threshold.
Since varying the \ac{EOS} mainly impacts the tidal deformabilities, which have a negligible effect on \ac{SNR}, we neglect selection effects in this study.

In practice, we first analyze individual events separately with Bayesian inference using an agnostic prior $\pi(\theta_i)$, giving single-event posteriors $p(\theta_i | d_i)$. 
Hence, we can write
\begin{equation}
    \mathcal{L}(\mathcal{D}|\thetaEOS) \propto \prod_{i = 1}^N \int \frac{p(\theta_i | d_i)}{\pi(\theta_i)} p(\theta_i | \thetaEOS) \diff \theta_i \, .
\end{equation}
Only the source-frame component masses $m_{1,2}$ and tidal deformabilities $\Lambda_{1,2}$ depend on $\thetaEOS$, so the above expression becomes
\begin{equation}
    \mathcal{L}(\mathcal{D}|\thetaEOS) \propto \prod_{i = 1}^N \int \frac{p(m_1, m_2, \Lambda_1, \Lambda_2 | d_i)}{\pi(m_1, m_2, \Lambda_1, \Lambda_2)} p(m_1, m_2, \Lambda_1, \Lambda_2 | \thetaEOS) \diff m_1 \diff m_2 \diff \Lambda_1 \diff \Lambda_2 \, .
\end{equation}
We use independent priors between $m_i$ and $\Lambda_i$, and uniform priors on $\Lambda_i$.
Moreover, since a sample $\thetaEOS$ corresponds to a deterministic function $\Lambda_{\mathrm{EOS}}(m)$, we find
\begin{equation}
    p(\Lambda_1, \Lambda_2 | m_1, m_2, \thetaEOS) = \delta\left(\Lambda_1 - \Lambda_{\mathrm{EOS}}(m_1)\right) \delta\left(\Lambda_2 - \Lambda_{\mathrm{EOS}}(m_2)\right) \, .
\end{equation}
Assuming a fixed uniform mass distribution, we then find
\begin{equation}
    \mathcal{L}(\mathcal{D}|\thetaEOS) \propto \prod_{i = 1}^N \int \frac{p(m_1, m_2, \Lambda_{\mathrm{EOS}}(m_1), \Lambda_{\mathrm{EOS}}(m_2) | d_i)}{\pi(m_1, m_2)} \diff m_1 \diff m_2 \, .
\end{equation}
The remaining integrals are approximated with Monte Carlo integration, using a \ac{NF} $q_i$ to approximate the conditional posterior density
\begin{equation}\label{eq:hierarchical-inference-trained-flow}
    q_i \approx p(\Lambda_1, \Lambda_2 | m_1, m_2, d_i) \, .
\end{equation}
Moreover, since the priors on the component masses do not vary significantly across the posterior samples, we approximate the hyperlikelihood by 
\begin{equation}
    \mathcal{L}(\mathcal{D}|\thetaEOS) \propto \prod_{i=1}^N \sum_{k = 1}^M q_i(\Lambda_{\mathrm{EOS}}(m_1^k), \Lambda_{\mathrm{EOS}}(m_2^k)| m_1^k, m_2^k)  \, , \quad m_1^k, m_2^k \sim p(m_1, m_2 | d_i) \, .
\end{equation}
The flows are trained with \textsc{flowjax}~\cite{ward2023flowjax} and are evaluated in single precision to reduce their computational cost. 
We use $M=200$ samples per Monte Carlo integration. 
In the future, we will more accurately incorporate the priors on the masses and selection effects, building on our recent work in Ref.~\cite{Koehn:2026nin}. 
This is necessary to avoid biases when considering non-trivial populations and cosmology~\cite{Wysocki:2020myz,Vitale:2020aaz,Golomb:2021tll,Essick:2023upv,Kedia:2024hvw,Ghosh:2024cwc,Golomb:2024lds,Toubiana:2025syw}.

For the \ac{EOS} parametrization, we use the model from Ref.~\cite{Koehn:2024set}, consisting of three parts.
First, at the lowest densities up to $0.5 \ \nsat$, we fix the \ac{EOS} to the crust from Ref.~\cite{Douchin:2001sv}.
At higher densities, we employ an \ac{EOS} of nucleonic matter in $\beta$-equilibrium parametrized by the \ac{MM}~\cite{Margueron:2017eqc,Margueron:2017lup}, which expands the energy per nucleon as a Taylor series.
Here, we truncate the series at fourth order.
However, since the above description might not hold throughout the entire sequence of \ac{NS} solutions, we switch to an agnostic speed-of-sound parametrization~\cite{Tews:2018iwm} at densities above $\nbreak$ (sampled in the range $[1, 2]~\nsat$). 
That is, we place $8$ grid points in the speed-of-sound plane $c_s^2(n)$, and linearly interpolate between them to obtain the speed-of-sound profile, which can be integrated to yield the \ac{EOS}~\cite{Somasundaram:2021clp}.
In total, our \ac{EOS} parametrization has $D = 26$ dimensions.
The priors are identical to those used in Ref.~\cite{Wouters:2025zju}.

\section{Sequential Monte Carlo}

We sample the above hyperposterior using \ac{SMC}, a method that samples from the posterior and estimates the Bayesian evidence simultaneously.
The sampling iteratively transforms a set of particles through a sequence of distributions $p_1, \dots, p_T$~\cite{del2006sequential,chopin2020introduction,dai2022invitation}.
Each $p_t$ is normalized and we write $p_t = \gamma_t / \mathcal{Z}_t$ for $t = 1, \dots, T$. 
For Bayesian inference, one typically starts from the prior $p_1 = \pi$ and terminates at the posterior $p_T = \mathcal{P}$. 
Let us first review the generic idea behind \ac{SMC}, after which we will discuss the two specific versions used in this work. 

At iteration $t$, a set of $N$ particles $\theta_{t-1}^i$, distributed according to $p_{t-1}$, is pushed to the next density $p_t$ by the following operations. 
First, the particles get assigned an importance weight 
\begin{equation}
    W_t^i = \frac{p_{t}(\theta^i_{t-1})}{p_{t-1}(\theta^i_{t-1})} = \frac{\mathcal{Z}_{t-1}}{\mathcal{Z}_t} w_t^i \, ,
\end{equation}
where the unnormalized weights are given by
\begin{equation}\label{eq:smc-weighting-function-general}
    w_t^i = \frac{\gamma_t(\theta^i_{t-1})}{\gamma_{t-1}(\theta^i_{t-1})} \, .
\end{equation}
If there is sufficient overlap between the two densities, we can assume the weighted set of particles is distributed according to $p_t$, and we can estimate the unknown ratio of evidences as
\begin{equation}\label{eq:smc-evidence-tracker}
    \frac{\mathcal{Z}_{t}}{\mathcal{Z}_{t-1}} \approx \frac1N \sum_{i=1}^N w_t^i \, ,
\end{equation}
which allows us to estimate the Bayesian evidence if we start from a proper prior.
Second, the particles are resampled according to their importance weights, giving a set of $N$ unweighted particles possibly containing duplicates. 
Finally, to avoid weight collapse after subsequent reweightings, we diversify the particles by moving them through an \ac{MCMC} kernel $\mathcal{K}_t$, having $p_t$ as its invariant density.
This step is often called the rejuvenation step.

This generic procedure of \ac{SMC} does not specify the sequence of distributions that should be traversed.
Below, we discuss two paths, referred to as likelihood tempering and data tempering. 

\subsection{Likelihood tempering}

One option, called likelihood tempering, constructs a geometric sequence, raising the likelihood to an inverse temperature parameter $\beta_t$:
\begin{equation}
    p_t(\theta) = \frac{\mathcal{L}(\theta)^{\beta_t}\pi(\theta)}{\mathcal{Z}_t} \, ,
\end{equation}
where $\beta_t \in [0, 1]$ goes from the prior ($\beta_1 = 0$) to the posterior ($\beta_T = 1$).
With this choice, the reweighting function in Eq.~\eqref{eq:smc-weighting-function-general} takes the form
\begin{equation}
    w_t^i = \mathcal{L}(\theta_{t-1}^i)^{\beta_t - \beta_{t-1}} \, .
\end{equation}
The temperature ladder can be set adaptively by requiring that the reweighting step preserves a certain \ac{ESS}~\cite{kish_ess}, which can be computed from the unnormalized weights with
\begin{equation}
    \mathrm{ESS}_t = \frac{(\sum_i w_t^i)^2}{\sum_i \left(w_t^i\right)^2} \, .
\end{equation}
By requiring that the next reweighting step maintains at least $\alpha N$ effective samples, with $\alpha \in [0,1]$ a hyperparameter set by the user, the temperature can be determined by solving
\begin{equation}\label{eq:smc-likelihood-tempering-adaptive-temperature}
    \beta_t = \sup_{\beta_t \in [\beta_{t-1}, 1]} \left\{ \mathrm{ESS}_t \geq \alpha N \right\} \, ,
\end{equation}
using, e.g., the bisection algorithm. 
The pseudocode for this version of \ac{SMC}, adapted from Ref.~\cite{karamanis2025persistent}, is given in Algorithm~\ref{alg:smc-likelihood-tempering}. 
The output of the algorithm is a set of particles $\{\theta_T^i\}_{i=1}^N$ distributed according to the posterior, and an estimate $\hat{\mathcal{Z}}_T$ of the Bayesian evidence.
Since we will use likelihood tempering as a substep in the next subsection, we will denote it by 
\begin{equation}
    \{\theta_T^i\}_{i=1}^N, \hat{\mathcal{Z}}_T = \texttt{SMC}_{\texttt{LT}}\left(\pi, \mathcal{L}\right) \, ,
\end{equation}
where we suppress the dependence on the hyperparameters and Markov kernels for simplicity.

We run \ac{SMC} with likelihood tempering as implemented in \textsc{blackjax}, setting $N = 2000$ and $\alpha = 0.9$.
We use a Gaussian random walk kernel for the rejuvenation step, with the covariance matrix tuned on the fly using the empirical covariance of the particles to target an optimal Metropolis-Hastings acceptance rate of $0.234$~\cite{neal2011optimal}, taking $20$ \ac{MCMC} steps and retaining only the particles from the last iteration. 

\begin{algorithm}
    \caption{\ac{SMC} with adaptive likelihood tempering}\label{alg:smc-likelihood-tempering}
    \KwData{Prior $\pi(\theta)$, likelihood $\mathcal{L}(\theta)$, number of particles $N$, target \ac{ESS} fraction $\alpha$, Markov kernels $\mathcal{K}_t$}
    \KwResult{Samples $\{\theta_T^i\}_{i=1}^N$ approximating the posterior $\mathcal{P}(\theta)$, and evidence estimate $\hat{\mathcal{Z}}_T$}
    $t \gets 1$, $\beta_1 \gets 0$, $\hat{\mathcal{Z}}_1 \gets 1$\;
    \For{$i \gets 1$ \KwTo $N$}{
        $\theta_1^i \sim \pi(\theta)$ \Comment*[r]{Sample from the prior}
    }
    \While{$\beta_t < 1$}{
        $t \gets t+1$\;
        $\beta_t \gets \sup_{\beta_t \in [\beta_{t-1}, 1]} \left\{ \mathrm{ESS}_t \geq \alpha N \right\}$ \Comment*[r]{Compute next temperature}
        \For{$i \gets 1$ \KwTo $N$}{
            $w_t^i \gets \mathcal{L}(\theta_{t-1}^i)^{\beta_t - \beta_{t-1}}$ \Comment*[r]{Reweight}
            $W_t^i \gets w_t^i \big/ \sum_{j=1}^N w_t^j$  \;
        }
        $\hat{\mathcal{Z}}_t \gets \hat{\mathcal{Z}}_{t-1} \times N^{-1} \sum_{i=1}^N w_t^i$ \Comment*[r]{Update evidence estimate}
        $\{\tilde\theta_t^i\}_{i=1}^N \sim \texttt{resample}\left(\{\theta_{t-1}^i\}_{i=1}^N, \{W_t^i\}_{i=1}^N\right)$ \Comment*[r]{Resample}
        \For{$i \gets 1$ \KwTo $N$}{
            $\theta_t^i \gets \mathcal{K}_t(\tilde\theta_t^i)$ \Comment*[r]{MCMC move}
        }
    }
\end{algorithm}

\subsection{Data tempering}

While likelihood tempering can be used for any Bayesian inference problem tackled by \ac{SMC}, here we also consider another choice for the sequence of distributions for hierarchical inference with growing datasets. 
In this scenario, one such choice for \ac{SMC} is to perform data tempering, also known as \ac{IBIS}~\cite{chopin2020introduction,chopin2002sequential}. 

Let us first consider the naive case where one data point is added in each iteration, such that the timestep $t$ coincides exactly with the data point $d_t$ and we have a dataset $\{d_1,\dots, d_T\}$. 
In this case, one can consider the path of partial posteriors~\cite{dai2022invitation}, i.e., where the intermediate density at iteration $t$ is given by
\begin{equation}
    \mathcal{P}_t \equiv \frac{\mathcal{L}(d_{1:t} | \theta) \pi(\theta)}{\mathcal{Z}_t} \, ,
\end{equation}
where we introduced the shorthand $d_{a:b} = \{d_a, \dots, d_b\}$. 
In that case, the importance weight from Eq.~\eqref{eq:smc-weighting-function-general} takes the form
\begin{equation}
    w_t^i = \mathcal{L}_t(\theta_{t-1}^i) \, , 
\end{equation}
where $\mathcal{L}_t(\theta) = \mathcal{L}(d_t | \theta)$ is the single-event likelihood.

This procedure results in a fixed path traversing the $T$ partial posteriors. 
In practice, however, it is not guaranteed that $\mathcal{P}_{t-1}$ and $\mathcal{P}_t$ are sufficiently similar, a requirement that is enforced by design in adaptive likelihood tempering. 
As a result, adding a single informative data point (e.g., a high-SNR GW event) might cause a sudden drop in the \ac{ESS} when transitioning between densities. 
Therefore, better performance can be achieved by a hybrid version of \ac{IBIS} that includes likelihood tempering to bridge between two partial posteriors~\cite{chopin2020introduction}. 
We refer to this as an `update step', which updates the set of particles that approximate our current posterior with new information. 
However, triggering this update step for each data point can become wasteful. 
For instance, when confronting a set of particles already conditioned on a large history of loud events, a quiet GW source barely changes the posterior, so the particles stay sufficiently diverse and no update is needed.
In short, we need an adaptive version of \ac{IBIS} that automatically determines when to run the update step.

Therefore, we generalize the above to process, at once, a batch of several data points at iteration $t$, rather than a single data point. 
We denote the index of the last data point of the batch by $b_t$ and introduce the notation
\begin{equation}
    \mathcal{L}_{a:b}(\theta) = \prod_{i = a + 1}^b \mathcal{L}_i(\theta) \, ,
\end{equation}
to refer to the batched likelihood for $\{d_{a+1}, \dots, d_b\}$. 
With this notation, the weight function for the batch becomes
\begin{equation}\label{eq:smc-data-tempering-weights-product}
    w_t^i(\theta_{t-1}^i) = \frac{\mathcal{L}_{1: b_t}(\theta_{t-1}^i)}{\mathcal{L}_{1: b_{t - 1}}(\theta_{t-1}^i)} = \mathcal{L}_{b_{t-1} : b_t}(\theta_{t-1}^i) \, .
\end{equation}
The $\mathrm{ESS}_t$ of this weighted set again allows us to adaptively determine the updating schedule. 
Similar to Eq.~\eqref{eq:smc-likelihood-tempering-adaptive-temperature}, we set
\begin{equation}\label{eq:smc-data-tempering-adaptive-batch}
    b_t = \sup_{b_t \in [b_{t-1}, T]} \{\mathrm{ESS}_t \geq \tilde{\alpha} N \} \, ,
\end{equation}
where $\tilde{\alpha} \in [0, 1]$ is another hyperparameter.
(Note that the main text calls this $\alpha$ for simplicity, and it should not be confused with the threshold $\alpha$ used in likelihood tempering, although they play similar roles.)

To turn the above into an algorithm, it is more natural to write $b_{t-1} = n$ and $b_t = n + q$, following the visualization from Fig.~\ref{fig:diagram} (i.e., the particles $\theta_{t-1}^i$ at time $t$ are distributed according to the partial posterior conditioned on the first $n$ events).
When processing a new event, if $\mathrm{ESS}_t$ remains above the threshold $\tilde{\alpha} N$, the event is simply added to a `queue', and we proceed to the next iteration. 
When, after $q$ events, the batch likelihood is sufficiently informative and $\mathrm{ESS}_t$ drops below the threshold, we update the posterior using \ac{SMC} with likelihood tempering. 
We can then interpret the partial posterior of iteration $t-1$ as the prior, while we anneal in the batch likelihood $\mathcal{L}_{n:n+q}$ toward the partial posterior conditioned on all $n+q$ events observed so far. 
Indeed, note that
\begin{equation}
    \mathcal{P}_{n+q}(\theta) = \frac{1}{\mathcal{Z}_{n+q}} \pi(\theta) \prod_{i=1}^{n+q}\mathcal{L}_{i}(\theta) = \frac{\mathcal{Z}_{n}}{\mathcal{Z}_{n+q}} \mathcal{P}_n(\theta) \mathcal{L}_{n:n+q}(\theta) \, ,
\end{equation}
and therefore we can think of $\mathcal{P}_n(\theta)$ as the prior, and $\mathcal{L}_{n:n+q}(\theta)$ as the likelihood to anneal in during \ac{SMC} likelihood tempering. 
Although $\mathcal{P}_n$ is not tractable, we only require a set of samples and its normalization constant to initialize Algorithm~\ref{alg:smc-likelihood-tempering}, both of which are provided by recursively applying the above data tempering algorithm.
Therefore, the initialization step in Algorithm~\ref{alg:smc-likelihood-tempering} is skipped, and instead the initial samples are set to the \ac{SMC} particles obtained from the previous update step, which are already distributed according to the `prior' $\mathcal{P}_n$.

The pseudocode of this hybrid data tempering algorithm is given in Algorithm~\ref{alg:smc-data-tempering}.
We keep the same hyperparameters for the inner \ac{SMC} likelihood tempering step, and set $\tilde{\alpha} = 0.5$. 
Note that, since the likelihoods factorize, the weights in Eq.~\eqref{eq:smc-data-tempering-weights-product} can be updated iteratively, only requiring a cheap reweighting operation at a cost of mere seconds per event. 
Similarly, we can also cheaply resample the particles to obtain samples from the partial posterior $\mathcal{P}_t$ and compute the evidence after each single event, with minimal extra cost.
However, for simplicity, here we only save the partial posteriors after update steps as intermediate results.

\begin{algorithm}
    \caption{\ac{SMC} with adaptive data tempering}\label{alg:smc-data-tempering}
    \KwData{Prior $\pi(\theta)$, single-event likelihoods $\mathcal{L}_t(\theta)$, number of particles $N$, \ac{ESS} thresholds $\alpha, \tilde{\alpha} \in [0, 1]$, Markov kernels $\mathcal{K}_t$}
    \KwResult{Samples $\{\theta_t^i\}_{i=1}^N$ approximating the partial posterior $\mathcal{P}_t(\theta)$, and evidence estimates $\hat{\mathcal{Z}}_t$}
    $t \gets 0$, $n \gets 0$, $q \gets 0$, $\hat{\mathcal{Z}}_0 \gets 1$\;
    \For{$i \gets 1$ \KwTo $N$}{
        $\theta_0^i \sim \pi(\theta)$ \Comment*[r]{Sample from the prior}
        $w_0^i \gets 1/N$ \Comment*[r]{Initialize weights}
    }
    \While{$t < T$}{
        $t \gets t+1$, $q \gets q+1$\;
        \For{$i \gets 1$ \KwTo $N$}{
            $w_t^i \gets w_{t-1}^i \times \mathcal{L}_t(\theta_n^i)$ \Comment*[r]{Reweight with data point t}
        }
        \If{$\mathrm{ESS}_t < \tilde{\alpha} N$ or $t = T$}{
            $\{\theta_t^i\}_{i=1}^N, \hat{\mathcal{Z}}_{n:n+q} \gets \texttt{SMC}_{\texttt{LT}}\left(\mathcal{P}_n, \mathcal{L}_{n:n+q} \right)$ \Comment*[r]{Alg.~\ref{alg:smc-likelihood-tempering}}
            $\hat{\mathcal{Z}}_t \gets \hat{\mathcal{Z}}_{n} \times \hat{\mathcal{Z}}_{n:n+q}$ \Comment*[r]{Update evidence estimate}
            $n \gets n+q$ \Comment*[r]{Update event count}
            $q \gets 0$ \Comment*[r]{Clear queue}
            $w_t^i \gets 1/N$ \Comment*[r]{Reset weights}
        }
    }
\end{algorithm}

\section{Mock Einstein Telescope dataset}\label{sec:gw-injections}

\subsection{Determining the fiducial EOS}\label{sec:injection-eos-generation}

To compute the tidal deformabilities of the injections, we determine a fiducial \ac{EOS} using the following \ac{NS} observations. 
First, we use the mass measurement of PSR~J0740+6620 of $(2.08 \pm 0.07)~\Msun$~\cite{Fonseca:2021wxt}.
Second, we use the NICER mass-radius measurements of PSR~J0030+0451~\cite{Riley:2019yda,Miller:2019cac}, PSR~J0740+6620~\cite{Miller:2021qha,Salmi:2024aum}, PSR~J0437--4715~\cite{Choudhury:2024xbk}, and PSR~J0614--3329~\cite{Mauviard:2025dmd}.
The latest data releases of PSR~J0437--4715~\cite{Miller:2025qfq}, PSR~J0030+0451~\cite{Kini:2026rjx}, PSR~J0614--3329~\cite{Miller:2026vpr}, PSR~J1614--2230~\cite{Mauviard:2026gzc}, and PSR~J2124--3358~\cite{Gonzalez-Caniulef:2026wcm} were not available at the start of this project and are not used.
Finally, we use the posterior samples of GW170817 analyzed with the \texttt{IMRPhenomXP\_NRTidalv3} waveform model available from Ref.~\cite{Wouters:2025csq}. 
The data are analyzed with \textsc{jester}, using the \ac{SMC} sampler with likelihood tempering.
We take the posterior sample with the maximum-likelihood value as our \ac{EOS} for simulating the injections.

\begin{figure}[h]
     \centering
     \includegraphics[width=\textwidth]{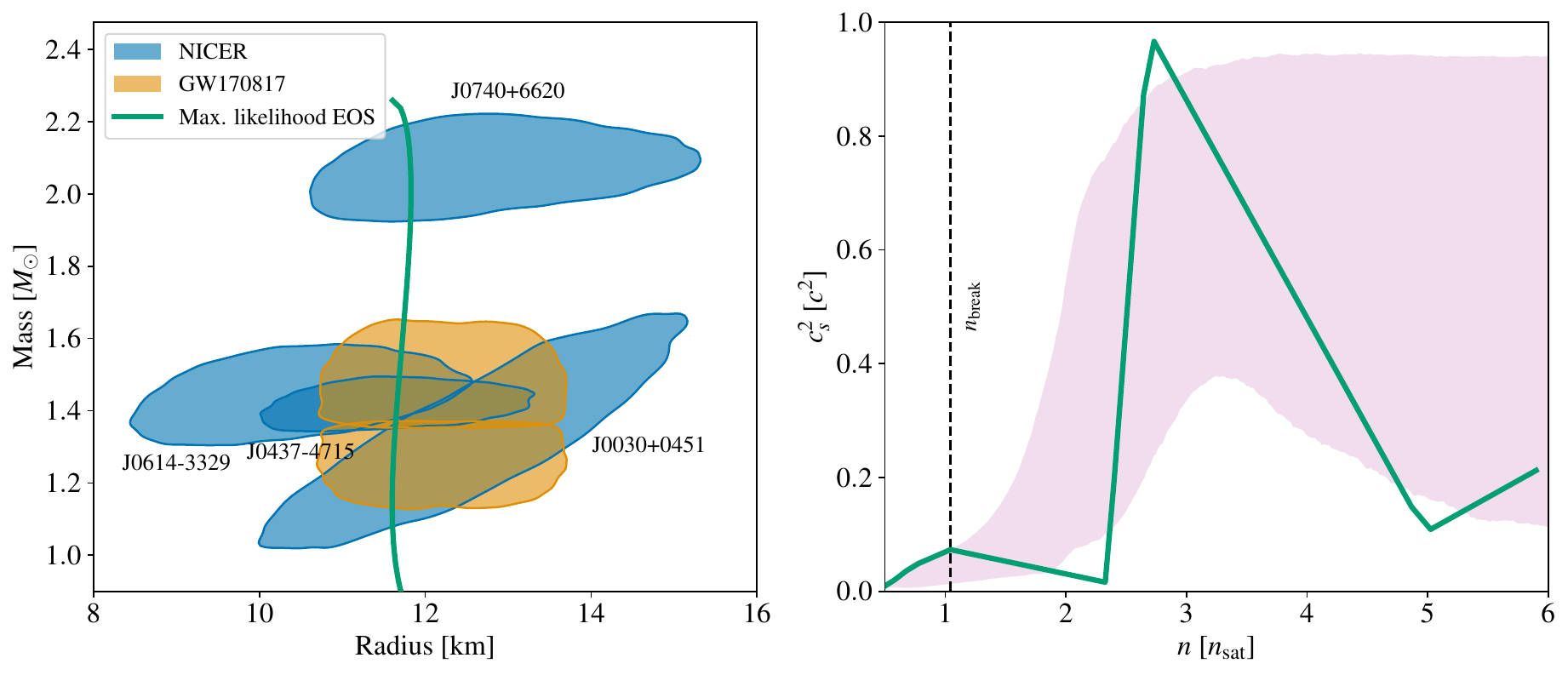}
     \caption{
        \textit{Left}: Constraints from NICER (blue) and GW170817 (orange) used to determine the maximum likelihood sample (green).
        \textit{Right}: Sound-speed profile of the maximum likelihood \ac{EOS} sample compared to the $95\%$ credible band of the posterior (pink).
    }
    \label{fig:real-data-run}
\end{figure}

Figure~\ref{fig:real-data-run} shows the mass-radius relation and sound-speed profile of this \ac{EOS}, together with the \ac{NS} data in the mass-radius plane. 
To be consistent with lower radii at intermediate masses as well as to support \acp{NS} with masses above $2~\Msun$, the maximum-likelihood \ac{EOS} remains at a low $c_s^2$ and suddenly stiffens around $2.5~\nsat$, after which it gradually decreases.
However, most samples in the posterior do not show a similar behavior, as demonstrated by the $95\%$ credible band.

\subsection{Single-event parameter estimation}\label{sec:single-event-pe}

Having chosen our injection \ac{EOS}, we generate a mock dataset of one month of \acp{BNS} observed with \ac{ET} using \textsc{gwmock}~\cite{gwmock_v0.13.2}.
We consider the $2$L configuration, with one L placed in the Meuse-Rhine euregion and the other in Sardinia, with the sensitivity curves taken from Ref.~\cite{Branchesi:2023mws}. 
We assume a local merger rate of ${106.6~\mathrm{Gpc}^{-3}~\mathrm{yr}^{-1}}$, draw source-frame masses uniformly between $[1, 2]~\Msun$ and aligned spins uniformly with dimensionless magnitude up to $0.05$.

Unfortunately, a complete Bayesian analysis of long-duration and loud signals observed with \ac{ET} still presents a significant computational bottleneck today~\cite{Couvares:2021ajn,Hu:2024mvn}.
To reduce computational cost as much as possible in the single-event parameter estimation, we use the \texttt{TaylorF2} waveform model~\cite{Buonanno:2009zt,Blanchet:2013haa} and analyze the signals from a lower frequency cutoff of $f_{\mathrm{min}} = 20~\mathrm{Hz}$.
However, we note that \ac{ET} is expected to be sensitive down to $5~\mathrm{Hz}$~\cite{ET:2025xjr}, which would increase the \ac{SNR} of the signals considered and lead to more detections (we provide the number of detected sources from $f_{\mathrm{min}} = 5~\mathrm{Hz}$ in the main text).
Such a low $f_{\mathrm{min}}$ increases the duration of the signal and makes Bayesian inference for a catalog this large infeasible with currently available methods.

In the end, our catalog contains $1579$ detected sources, applying an \ac{SNR} cutoff of $12$~\cite{KAGRA:2013rdx}.
A few long-duration events do not converge within the imposed time limits and are removed from our catalog, leaving $1555$ events.
We perform parameter estimation on the signals using nested sampling~\cite{Skilling:2004pqw,Skilling:2006gxv}, in particular with the \textsc{dynesty} sampler~\cite{Speagle:2019ivv,sergey_koposov_2026_21417233} with $2000$ live points and otherwise identical settings to its implementation in \textsc{bilby}~\cite{Ashton:2018jfp, Romero-Shaw:2020owr}.
To further reduce computational costs, we use the multibanding likelihood~\cite{Garcia-Quiros:2020qlt,Morisaki:2021ngj} and only sample the subspace of intrinsic parameters that is relevant for our \ac{EOS} inference, namely the detector-frame chirp mass $\mathcal{M}_c$, mass ratio $q$, and component tidal deformabilities $\Lambda_{1,2}$, with uniform priors.
Spins and extrinsic parameters are fixed to their injected values during the recovery.
Of the fixed parameters, fixing the luminosity distance $d_L$ can impact the \ac{EOS} inference, as it removes the uncertainty in the redshift, which is needed to transform detector-frame masses to source-frame masses.
In the current setup, the redshift is fixed by using the Planck2018 cosmology to convert distance into redshift. 
On the other hand, our mass posteriors are broader since we analyze signals from $20$ Hz instead of $5$ Hz.
With this setup, we achieve a median sampling runtime of around $10$ minutes per signal on $128$ CPUs, making the analysis of many signals manageable.
We analyze the signals both in stationary, Gaussian detector noise, as used for the results in the main text, and in zero noise, for which the \ac{EOS} recovery is shown in Fig.~\ref{fig:zero-noise-results} below.

As defined in Eq.~\eqref{eq:hierarchical-inference-trained-flow}, the single-event posteriors enter the hierarchical likelihood through an \ac{NF} that approximates the conditional posterior density $p(\Lambda_1, \Lambda_2 | m_1, m_2, d_i)$. 
This flow is trained at the end of each \textsc{bilby} run using \textsc{flowjax}~\cite{ward2023flowjax}. 
We use the masked autoregressive flow architecture with rational quadratic spline transforms~\cite{Papamakarios:2017tec,Durkan:2019nsq}. 
The training (performed on CPUs) finishes within $2$ minutes and therefore does not present a bottleneck in the analysis.

With a median runtime of about $12$ minutes for each of the $1555$ \textsc{bilby} runs (including flow training), the wall time is currently dominated by the analysis of individual events, not by the hierarchical \ac{EOS} inference, especially considering our simplified settings for the \textsc{bilby} runs.
In the future, we can therefore explore the use of recent advances in parameter estimation of long-duration signals~\cite{Smith:2021bqc,Dax:2024mcn,Hu:2024lrj,Tenorio:2025gci,Yallup:2026nqc,Guttman:2026fpc,Sengupta:2026hif,OShaughnessy:2026kkf} to sample more parameters, extend the analysis to lower frequencies, make the mock dataset more realistic, and include effects from Earth rotation~\cite{Virtuoso:2024kyp,Baral:2025geo} and overlapping signals~\cite{Samajdar:2021egv,Antonelli:2021vwg,Janquart:2022fzz}.

\section{Zero-noise injections}

\begin{figure}
     \centering
     \begin{minipage}[c]{0.495\textwidth}
         \centering
         \includegraphics[width=\textwidth]{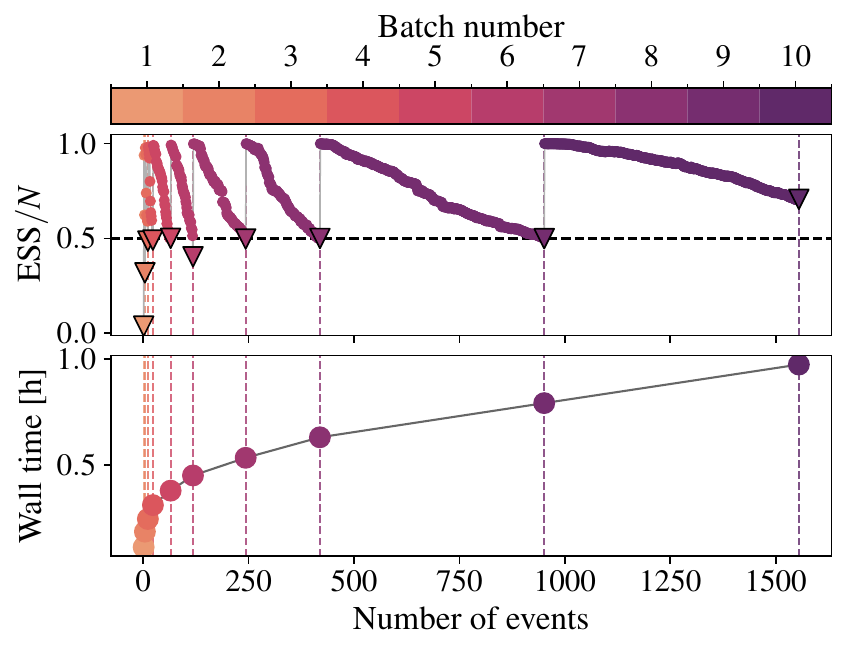}
     \end{minipage}
     \hfill
     \begin{minipage}[c]{0.495\textwidth}
         \centering
         \includegraphics[width=\textwidth]{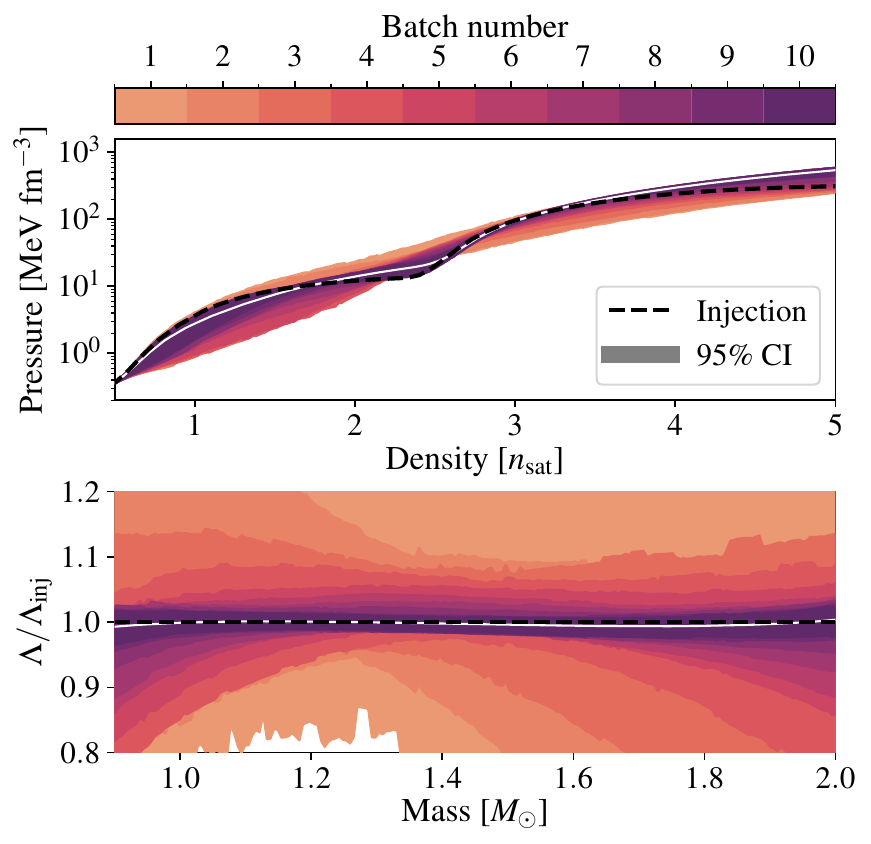}
     \end{minipage}
    \caption{
        Results for the \ac{SMC} diagnostics (\textit{left panels}) and the \ac{EOS} constraints (\textit{right panels}) when analyzing the catalog of one month of \ac{BNS} sources, simulated without detector noise.
        The \ac{ESS} decreases more smoothly, and the slight bias in the recovered mass--tidal-deformability curve observed in Fig.~\ref{fig:eos-constraints} is removed. 
        }
    \label{fig:zero-noise-results}
\end{figure}

With the setup described in the main text, we create two mock datasets: one in which Gaussian noise is added to the data, and one in which the signals are analyzed with zero noise. 
The injections with noise are presented as the main results in the main text. 
Figure~\ref{fig:zero-noise-results} shows the results of our \ac{SMC} sampler on the noiseless injections.
Without noise, the sampling efficiency decays gradually and monotonically as more events are added, rather than showing the erratic behavior observed when analyzing the noisy injections (Fig.~\ref{fig:smc-updating-diagnostics}).
As a result, only half as many batches are needed to process the events, and the total wall time is roughly halved as well.
Regarding the derived constraints, we notice that the median tidal deformability curve coincides much more closely with the true injected value than in the noisy injections.
Therefore, we conclude that the presence of noise, even when analyzing many events, can cause the recovered tidal deformability to deviate from the injected value.

\end{document}